\documentclass{article} 
\usepackage{iclr2020_conference,times}

\usepackage{amsmath,amsfonts,bm}

\def\eqref#1{equation~\ref{#1}}

\def\1{\bm{1}}

\DeclareMathAlphabet{\mathsfit}{\encodingdefault}{\sfdefault}{m}{sl}
\SetMathAlphabet{\mathsfit}{bold}{\encodingdefault}{\sfdefault}{bx}{n}

\usepackage{hyperref}
\usepackage{url}
\usepackage{graphicx}
\usepackage{xcolor}
\usepackage{natbib}
\usepackage{caption} 
\usepackage{gensymb}
\title{A single image deep learning approach to restoration of corrupted remote sensing products}

\iclrfinalcopy

\author{Anna Petrovskaia, Raghavendra B. Jana,  \& Ivan V. Oseledets \\
Center for Computational and Data-Intensive Science and Engineering\\
Skolkovo Institute of Science and Technology\\
Moscow, 121205, Russia \\
\texttt{\{anna.petrovskaia, r.jana, i.oseledets\}@skoltech.ru} \\
}

\begin{document}

\maketitle

\begin{abstract}
Remote sensing images are used for a variety of analyses, from agricultural monitoring, to disaster relief, to resource planning, among others. The images can be corrupted due to a number of reasons, including instrument errors and natural obstacles such as clouds. We present here a novel approach for reconstruction of missing information in such cases using only the corrupted image as the input. The Deep Image Prior methodology eliminates the need for a pre-trained network or an image database. It is shown that the approach easily beats the performance of traditional single-image methods.

\end{abstract}

\section{Introduction}
Remote sensing is increasingly recognised as a convenient tool with a wide variety of uses in agriculture, including but not limited to crop type identification and classification, crop-yield forecasting, crop condition assessment and stress detection, and land cover and land degradation mapping. In light of increased demand for remote sensing data, NASA's Landsat program has received considerable attention.
Landsat-7 has supplied multi-spectral imagery of the Earth’s surface for more than four years and has become an important data source for a large number of research and policy-making initiatives.
Unfortunately, a Scan Line Corrector (SLC), which compensates for the forward motion of the satellite, broke down in May 2003 \citep{usgs}. Since the failure of the SLC, Landsat-7 scenes exhibit wedge-shaped swaths of missing data that reach 14 pixels in width near the edges. It is estimated that up to 22 percent of any given scene could be lost because of the SLC failure, and makes a significant number of Landsat imagery unsuitable for further use in research devoted to any agricultural needs.

Numerous learning-free gap-filling approaches have been tried before to enable inference of the missing pixels (e.g. kriging and co-kriging \citep{chiles2009geostatistics, zhang2007gaps}, Geostatistical neighbourhood similar pixel interpolator \citep{zhu2012new}, Weighted Linear Regression \citep{zeng2013recovering}, Direct Sampling \citep{mariethoz2010direct}, Localised Linear Histogram Matching \citep{scaramuzza2004slc}). The approaches can be broadly classified into two types - single-image methods, and multi-image methods. Multi-image methods generally make use of non-corrupted imagery available from a temporally close overpass to fill in the gaps in the current (corrupted) image, and have been shown to be more accurate than the single-image approaches. The performance of the multi-image methods is inherently dependent on the interval between the acquisition of the two images - greater the interval, lower the accuracy of reconstruction, especially in fast-changing conditions such as agricultural regions. However, in many cases, it may not be possible to find a non-corrupted image within a time frame suitable for high-accuracy reconstruction. This is especially the situation in regions with a high degree of cloud cover, and very few clear days. In such cases, one has no other option but to fall back on single-image approaches. 

In this paper, we present one such single-image approach based on leveraging the abilities of the Deep Image Prior (DIP) method \citep{ulyanov2018deep} to fill in gaps using only the corrupt image. We also test the ability of DIP to reconstruct remote sensing scenes with different levels of corruption in them. Finally, we also compare the performance of our approach with the performance of classical single-image gap-filling methods.

\section{Methodology}

\subsection{Study area and data}

To investigate the accuracy of reconstruction by Deep Image Prior, we use a non-corrupted image from 24th of July, 2002 and a gap mask that mimics the SLC-off condition.
The area considered is rather heterogeneous, consisting of vegetation cover, bare soil, open water, and impervious surfaces. It is located in California, USA, around 37.97 \degree N, 121.51 \degree W and comprises an area of 12 $km^2$. The choice of our study area was influenced by the same image having been used in similar studies previously. Using the same image/area helps in comparing the performance of our method versus earlier approaches.
Gap masks of different widths were overlaid on the original image to simulate different percentages of corrupted pixels, thus simulating different parts of the Landsat scene that could be corrupted to different levels. 
For the comparison of gap-filling methods Landsat-7 bands 1-4 were used in the study. The bands correspond to the red, green, and blue portions of the visible spectrum and near-infrared spectral range. 

\subsection{Deep Image Prior}

“Deep Image Prior” (DIP) \citep{ulyanov2018deep}is an approach for training convolutional neural networks (CNN) \citep{lecun1989backpropagation} with image data. The DIP methodology eliminates the need for a pre-trained network or an image database. Only the corrupted image (designated as ${x}_{0}$) is used in the restoration process.

The DIP method is based on the assumption that the image prior can be found within a CNN itself and need not be learned from a separate training dataset or designed manually.
An optimization objective in image restoration tasks is often defined as 

\begingroup
\begin{equation}
    {\bf \min_{x} \textit E ({x}; {x}_{0}) + \textit {R}({x})}, 
\end{equation}
\endgroup

where ${\textbf {\textit x}}$ is the original image, ${\textbf{\textit x}_0}$ is the corrupted image,
$ \textit E ({x}; {x}_{0})$ is the data term which is negative log of the likelihood and $\textit R(x)$ is the image prior term which is negative log of the prior.

The usual approach is to initialize x with some random noise, compute the gradient of the function with respect to x, and traverse the image space until it converges at some point. So, the optimization is evaluated in the image space. Unlike the conventional approach, the DIP approach \cite{ulyanov2018deep} proposes to perform a surjective $g : \theta \mapsto x $. In this way, we obtain 

\begingroup
\begin{equation}
    {\bf  \min_{x} \textit E ({g}(\theta); {x}_{0}) + \textit {R}({g}(\theta))}. 
\end{equation}
\endgroup

This equation is, in theory, equivalent to (1). In this approach, the function ${\bf g}$ is initialized with random values of $\theta$. The output from function ${\bf g}$ is then mapped to the image space and updated $\theta$ using gradient descent.

The novel approach gets rid of the prior term by selecting appropriate ${\bf g}$.
${\bf g}(\theta)$ can be defined as ${\bf \textit f_{\theta}(z)}$, where ${\bf \textit f} $ is a deep convolutional network with parameters $\theta$ and ${\bf z}$ is a fixed input. The equation(2) is then reformulated as
\begingroup
\begin{equation}
    {\bf  \min_{\theta} \textit E ( \textit f_{\theta}(z); {x}_{0})}. 
\end{equation}
\endgroup

Thus, instead of searching for the answer in the image space we now search for it in the space of the neural network's parameters.

\subsection{Classical single-image gap-filling methods}

The performance of our DIP approach is compared with three other popular gap-filling methods: kriging interpolation \citep{chiles2009geostatistics}), weighted linear regression \citep{zeng2013recovering}  and direct sampling method \citep{mariethoz2010direct}. These methods used learning-free techniques, and hence, are single-image techniques, as is our DIP approach.
The simulation results of these methods were taken from the paper of \citet{yin2017comparison}. To make our results comparable with the numbers from the paper we used exactly the same data as in the above mentioned article.

\subsection{Experimental setting}

The satellite image restoration was carried out in two different ways. Firstly, we filled the gaps for each band separately. In this case an input $x_0$ for Deep Image Prior network was a single band. The number of hidden (or corrupted) pixels was fixed at 55\%. In the second approach, we stacked all four spectral bands and trained the network on the obtained composite. We used the composite to estimate the influence of the number of corrupted pixels on the ability of Deep Image Prior to fill values. We conducted 5 simulations with the image corruption levels of 3\%, 6\%, 15\%, 35\% and 55\% of the whole area. 

We evaluate the performance of the gap-filling simulations using classical metrics: root mean squared error(RMSE) and $r^2$ score. RMSE is a metric representing square root of average squared error computed for each pixel. Lower values of RMSE indicate more accurate simulations. However, RMSE could be a bad metric in the case of noisy data. $r^2$ score or the coefficient of determination provides information about the variance of the simulation reconstruction in comparison with the true value. $r^2$ shows the quality of model fitness, with higher values indicating better predictions.

 For our convolutional network,we use the “hourglass” (also known as “decoder-encoder”) architecture as that used in the original Deep Image Prior paper \cite{ulyanov2018deep}. The main part of code regarding the CNN training was acquired from the publicly available PyTorch implementation\footnote{https://github.com/DmitryUlyanov/deep-image-prior}. We trained our Deep Image Prior model for 1500 epochs using the Adam solver \citep{kingma2014adam} with a batch size of 1. We use LeakyReLU \citep{he2015delving} as a non-linearity. As an up-sampling operation the nearest neighbour up-sampling was used. The input vector $z$ was filled with uniform noise between zero and 0.1.

\section{Results}
\begin{figure}[h!]
\centering
\includegraphics[width=0.9\textwidth]{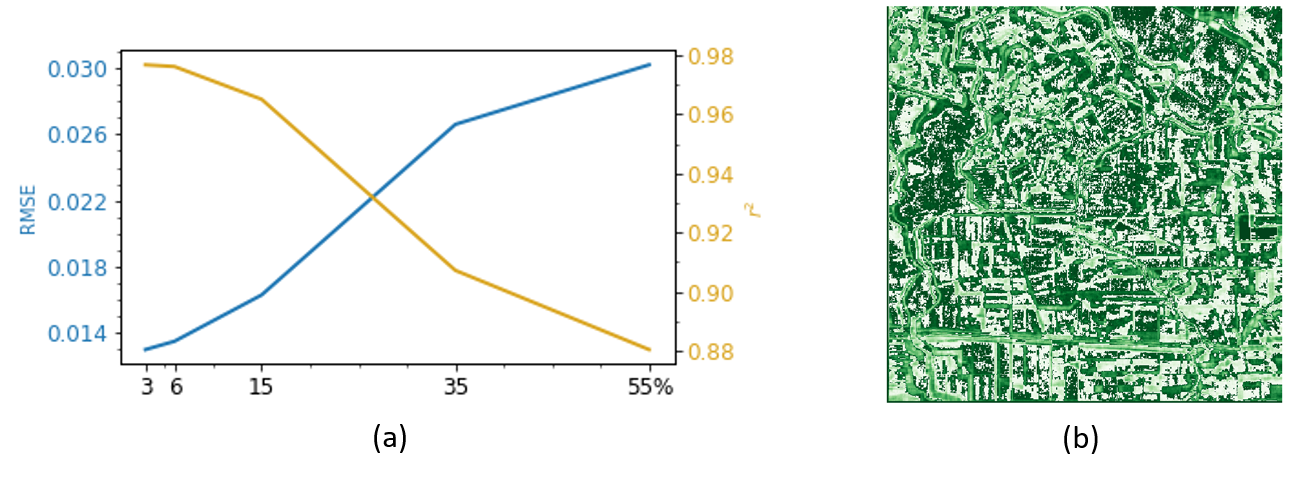}
\caption{Simulation results: (a) Comparative statistics; and (b) Pixel-wise similarity visualization for case with 55\% hidden pixels (worst case scenario). In (a), x-axis: Percentage of hidden/corrupt pixels, y-axis (left): root mean squared error (RMSE) between simulated and real pixel values, and y-axis (right): corresponding coefficient of determination ($r^2$). In (b), darker green shows higher level of similarity, while white represents dissimilarity.}
\label{fig:fig}
\end{figure}

The quantitative comparison on the separate bands of given data for our method is given in Table \ref{table}, showing a quantitative advantage of the proposed approach compared to classical gap-filling methods. $r^2$ is significantly higher for each reconstructed band. The lowest performing restoration made by DIP approach reaches 0.812 (Band 1), while the best value for classical approaches for this band is 0.685 achieved by the DS method. This result shows the capability of Deep Image Prior to fit the model. This result also demonstrates that Deep Image Prior is able to identify the patterns in the satellite image remarkably better than the other methods considered. However, as regards RMSE, DIP approach have noticeably higher values. The explanation for such a mismatch is connected to the process of DIP training and the procedure of filling the gaps. Strictly speaking, DIP does not simply fill in only the missing values. It reconstructs the entire image using the non-corrupted parts as guidelines from the training. This peculiarity of the training process allows DIP to achieve impressive results with regard to the correlation between the prediction and reality, but at the same time the non-corrupted parts of the initial image become slightly degraded after the restoration process. It can be clearly seen from Figure \ref{fig:fig} b which shows the similarity between original image (ground truth) and reconstruction from 55\% hidden pixels scenario. The figure is very heterogeneous, and the parts that were corrupted on the original image can not be distinguished from this figure. 

\begin{table}[ht]
\centering
\caption{Comparison of performance of different popular gap-filling methods \citep{yin2017comparison} and Deep Image Prior. Bold text highlights the best value for a band. DIP approach outperform the other methods compared by $r^2$ score for all bands. }
\label{table}
\begin{tabular}{|l|cccc|cccc|}
\hline
 & \multicolumn{4}{c|}{\bf RMSE} & \multicolumn{4}{c|}{$\bf r^2$} \\
Method & Band 1 & Band 2 & Band 3 & Band 4 & Band 1 & Band 2 & Band 3 & Band 4 \\ \hline
Kriging & 0.010 & 0.015  & 0.023 & 0.063 & 0.610 & 0.627  & 0.728 & 0.690   \\
WLR  & 0.010 & 0.014  & 0.023 & 0.055 & 0.622 & 0.694  & 0.742 & 0.765 \\
DS  & \bf 0.009 & \bf 0.012  & \bf 0.020 & \bf 0.052 & 0.685 & 0.755  & 0.792 & 0.780 \\
DIP (our) & 0.020 & 0.024 &  0.043 & \bf 0.052 & \bf 0.812  & \bf 0.853  & \bf 0.874 & \bf 0.832 \\ 
\hline
\end{tabular}
\end{table}

As for training settings, we found that using a composite of four bands gives better results than reconstructing each band separately. We conducted this experiment with 55\% of corrupted values - the highest possible level of corruption in the Landsat 7 scenes. The average $r^2$ for separate training is 0.842, $r^2$ for composite - 0.880. The performance with respect to RMSE was also marginally better for composite training than for separate bands: 0.030 versus 0.034 respectively. 

To examine the robustness of DIP, we compared the influence of the number of corrupted pixels on the restoration results. As can be seen in Figure \ref{fig:fig} (a), the Deep Image Prior is able to successfully handle both small and large number of hidden pixels.

\begin{figure}[h!]
\centering
\includegraphics[width=1\textwidth]{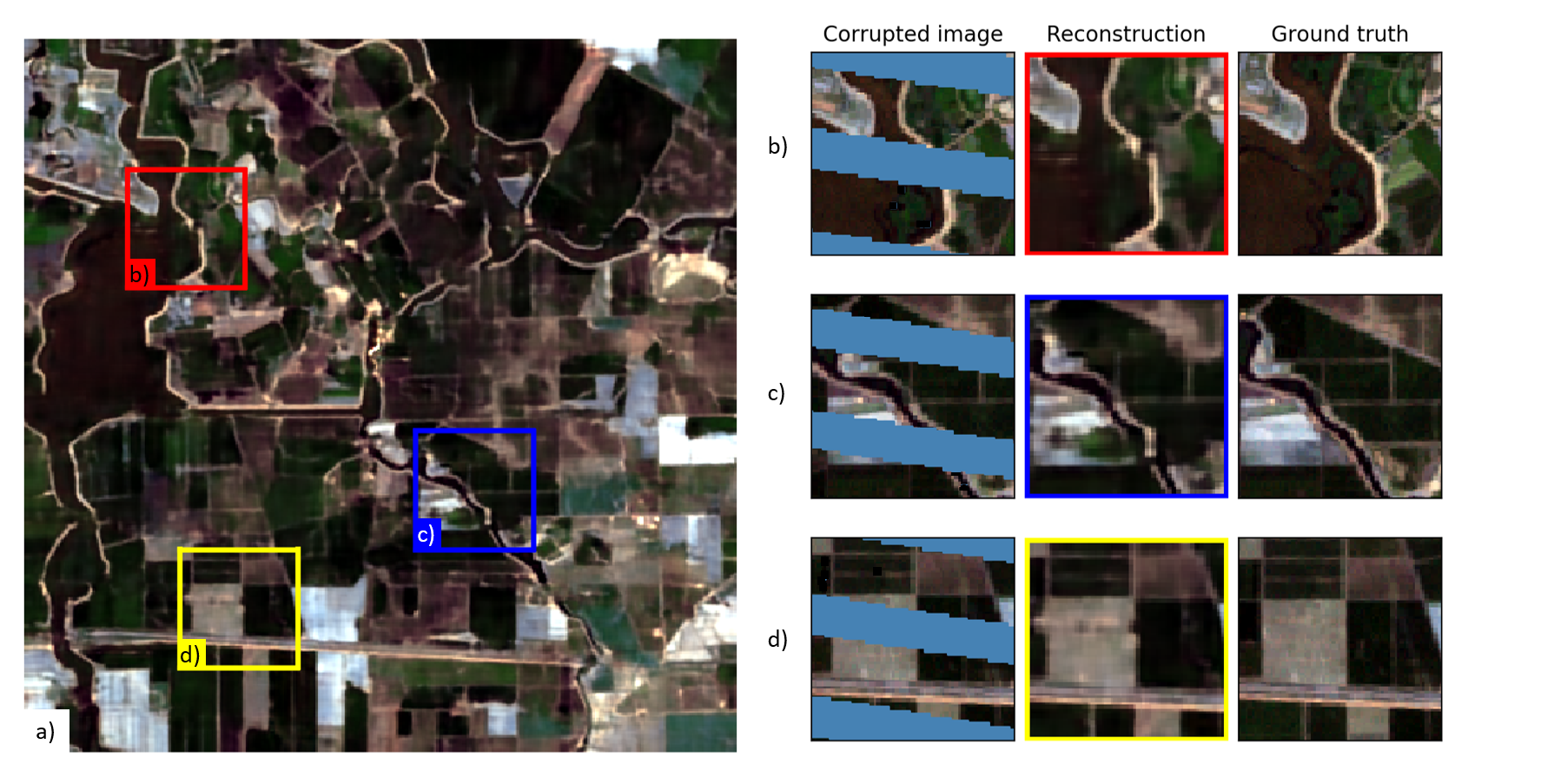}
\caption{Comparison of reconstruction (55\% hidden pixels), image with gap mask and original image for three regions. (a) - result of reconstruction, (b),(c),(d) - parts of image overlapped with gap mask (corrupted image), corresponding parts of reconstructed image and parts of initial image (ground truth).}
\label{fig:reconstruction}
\end{figure}

In Figure \ref{fig:reconstruction} we present a qualitative visual comparison of the reconstructed image with the original image and the image covered by gap mask. This comparison is made on the most difficult case with the largest amount of hidden pixels (55\%). Analysing particular parts of image we found the gap filling to be quite accurate except for minor inaccuracies.

\section{Conclusion}
This work sets an example for applying a deep learning approach to the restoration of corrupted remote sensing imagery where multi-temporal snapshots may be unavailable due to reasons such as cloud cover or instrument failure. We have successfully demonstrated the superior capability of our approach over traditional single-image methods in learning the spatial patterns in the image. This approach could be further extended to problems with irregular artifacts such as the removal of clouds, which could further enhance the usability of remote sensing imagery. The usage of this approach will expand the possibilities for a wide variety of agricultural studies and applications.


\bibliography{iclr2020_conference}
\bibliographystyle{iclr2020_conference}

\appendix
\section{Appendix}
\subsection{Architecture details}
\begin{figure}[h!]
\centering
\includegraphics[width=1\textwidth]{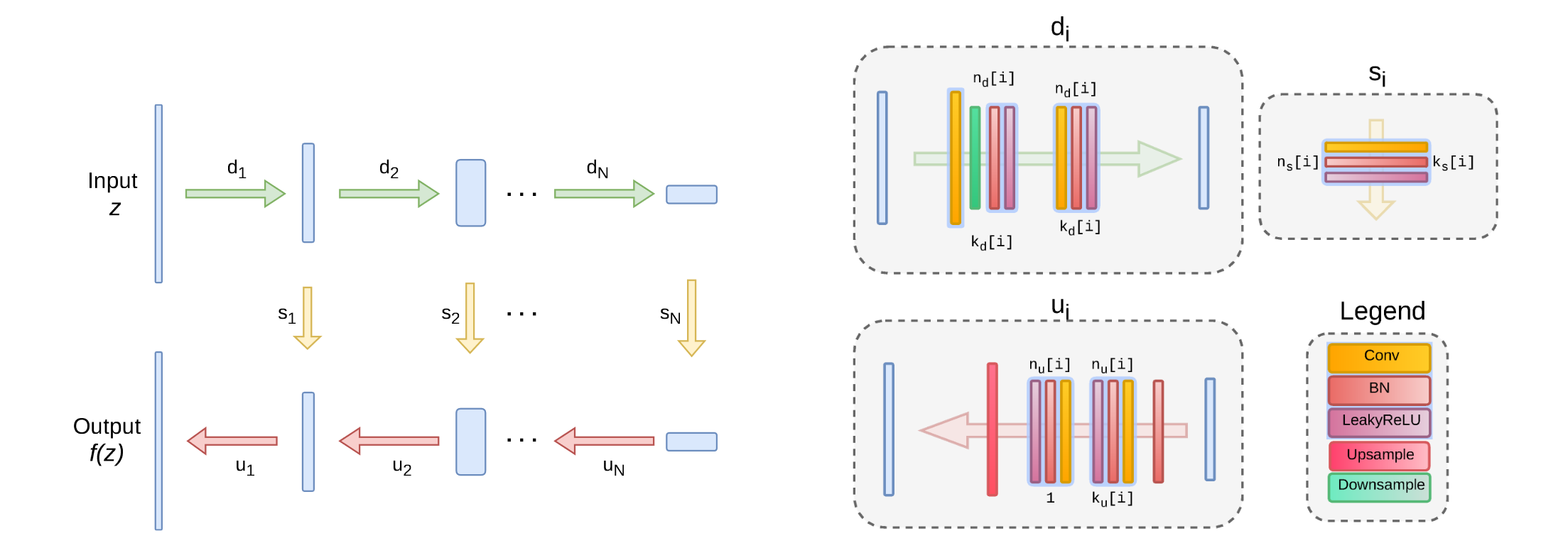}
\caption{The architechture used for image restoration. $n_{u}[i], n_{d}[i], n_{s}[i]$ correspond to the number of filters at depth $i$ for the upsampling, downsampling and skip-connections respectively. The values $k_{u}[i], k_{d}[i], k_{s}[i]$ correspond to the respective kernel sizes. The image taken from the suplemantary materials of the paper \citep{ulyanov2018deep}.
}
\label{fig:arch}
\end{figure}

The architecture and hyperparameters were chosen following the paper \citep{ulyanov2018deep}.
We used the encoder-decoder architecture with 4 skip-connections layers. As a down-sampling technique, we used  strides  implemented
within  convolution  modules.  As an up-sampling operation, we used nearest neighbour up-sampling. The input for the network was meshgrid, initialized with $z \in \mathbb{R}^{2 \times W \times H}$
using \texttt{np.meshgrid}. The other details of the architecture are provided below. The explanation is on the figure \ref{fig:arch} taken from the supplementary materials of the paper \citep{ulyanov2018deep}.

\begin{tabular}{|l|}
\hline
\\
$z \in \mathbb{R}^{4 \times W \times H} \sim U\left(0, \frac{1}{10}\right)$\\
$n_{u}=n_{d}$\texttt{ =[128,128,128,128,128]}\\
$k_{u}=k_{d}=$ \texttt{[3,3,3,3,3]}\\
$n_{s}=$\texttt{[128,128,128,128,128]}\\
$k_{s}=$\texttt{[1,1,1,1,1]} \\
$\sigma_{p}=\frac{1}{10}$ \\
\texttt{num\_iter =1500}\\
\texttt{LR=0.01} \\
\texttt{upsampling = nearest}\\
\\
\hline
\end{tabular}

\end{document}